\documentclass[11pt, a4paper]{article}

\usepackage{jheppub}
\usepackage{amsmath, bm, braket, orcidlink}

\begin{document}


\title{
Implications of tachyonic phase transition in classically scale invariant general $U(1)_{X}$ models 
}


\author[a,b]{Arindam Das\,\orcidlink{0000-0002-8934-2300}}
\emailAdd{adas@particle.sci.hokudai.ac.jp}
\affiliation[a]{Institute for Academic Innovation, Hokkaido University,
Sapporo, 060-0817, Japan}
\affiliation[b]{Department of Physics, Hokkaido University, Sapporo 060-0810, Japan}  

\author[c]{Katsuya Hashino\,\orcidlink{0000-0002-8459-9430}}
\emailAdd{hashino@fukushima-nct.ac.jp}
\affiliation[c]{National Institute of Technology, Fukushima College, Nagao 30, Taira-Kamiarakawa, Iwaki, Fukushima 970–8034, Japan}

\author[d]{Yuta Orikasa\,\orcidlink{0000-0003-0112-8682}}
\emailAdd{yuta.orikasa@utef.cvut.cz}
\affiliation[d]{Institute of Experimental and Applied Physics, Czech Technical University in Prague, Husova 240/5, 110 00 Prague 1, Czech Republic}

\author[e]{Masanori Tanaka\,\orcidlink{0000-0002-1303-7043}}
\emailAdd{tanaka@pku.edu.cn}
\affiliation[e]{Center for High Energy Physics, Peking University, Beijing 100871, China}

\abstract{
We investigate the complementarity between collider searches and gravitational-wave (GW) observations in the classically scale-invariant general $U(1)_X$ extension of the Standard Model. The $U(1)_X$ scale is generated radiatively through the Coleman-Weinberg mechanism, while the electroweak scale is induced through the Higgs portal, without explicit mass terms in the scalar potential. Three right-handed neutrinos are introduced to ensure anomaly cancellation and acquire Majorana masses upon $U(1)_X$ symmetry breaking, which also generates the mass of the neutral gauge boson $Z'$. In the strongly supercooled regime, a tachyonic $U(1)_X$ phase transition can produce a stochastic GW background. Assuming three heavy Majorana neutrinos, we study their effects on the phase-transition dynamics and reheating, and estimate the resulting GW signals. We identify the regions of the $U(1)_X$ gauge coupling and $Z'$ mass accessible to future GW observatories, including LISA and DECIGO, and compare them with existing LEP and LHC constraints. We find that GWs from tachyonic phase transitions can probe regions with small gauge couplings and heavy $Z'$ bosons that are difficult to access through collider searches, demonstrating the complementarity of these probes.
}

\maketitle

\section{Introduction}

The gauge hierarchy problem remains one of the most significant theoretical challenges in the Standard Model (SM), having driven the search for new physics for decades. 
This issue arises because quantum corrections to the Higgs doublet self-energy are quadratically divergent; if cut off by a physical scale far above the electroweak scale, maintaining the observed Higgs mass requires extreme fine-tuning of the tree-level parameter.
However, due to its chiral nature, the SM Lagrangian is classically scale-invariant (conformal), with the sole exception of the Higgs mass term.
As argued by Bardeen \cite{Bardeen:1995kv}, if this classical conformal symmetry—along with its minimal violation by quantum anomalies—is enforced, the SM can remain free of quadratic divergences, offering a potential solution to the hierarchy problem.
This framework aligns seamlessly with the scenario explored by Coleman and Weinberg \cite{Coleman:1973jx}, who demonstrated that a $U(1)$ gauge theory with a massless Higgs field can undergo radiative symmetry breaking via the effective potential.

Although applying the Coleman-Weinberg (CW) mechanism directly to the SM Higgs sector is appealing, it fails for the physical top-quark and weak-boson masses because the resulting effective potential is unbounded from below \cite{Politzer:1978ic, Fujikawa:1978ru}. Pursuing this framework requires extending the SM, a strategy adopted by numerous models (e.g., see Refs.~\cite{Hempfling:1996ht,Espinosa:2007qk,Chang:2007ki,Foot:2007ay,Foot:2007iy,
Iso:2009ss,Iso:2009nw,Das:2015nwk,Hashino:2015nxa,Oda:2015gna,Das:2016zue,Hashino:2016rvx,Khoze:2014xha,Braathen:2020vwo,Ahriche:2023jdq}). 
In this paper, we consider the classically conformal $U(1)_X$ extension proposed in Refs.~\cite{Oda:2015gna,Das:2016zue}, which supplements the SM content with three generations of right-handed neutrinos (RHNs) and a $U(1)_X$ Higgs field.
Requiring generation-independent fermion charges renders the model free from all gauge and gravitational anomalies while preserving the SM Yukawa structure, identifying the $U(1)_X$ symmetry as a linear combination of $U(1)_Y$ and $U(1)_{B-L}$ \cite{Appelquist:2002mw}. 
Our setup thus generalizes the classically conformal $U(1)_{B-L}$ extension studied in Refs.~\cite{Iso:2009ss,Iso:2009nw}.
Radiative symmetry breaking via the CW mechanism generates masses for both the $Z^\prime$ boson and the right-handed Majorana neutrinos. 
A portal quartic coupling between the $U(1)_X$ scalar and the SM Higgs doublet then induces a negative mass squared for the SM Higgs, driving electroweak (EW) symmetry breaking. 
Hence, the radiative breaking of the $U(1)_X$ symmetry serves as the sole origin of mass scales in this model. Finally, the heavy Majorana neutrinos naturally execute the seesaw mechanism \cite{Minkowski:1977sc,Yanagida:1979as,Gell-Mann:1979vob,Mohapatra:1979ia,Yanagida:1980xy,Schechter:1980gr}, generating tiny active neutrino masses and their flavor mixing following the EW symmetry breaking.

The phase transition dynamics for the $U(1)_{X}$ symmetry breaking have been thoroughly discussed~\cite{Iso:2017uuu,Chiang:2017zbz,Marzo:2018nov,Ellis:2019oqb,Hasegawa:2019amx,Bian:2019szo,Ellis:2020nnr,Borah:2021ocu,Badger:2022nwo,Kierkla:2022odc,Gouttenoire:2023pxh,deBoer:2024jne,Baratella:2024hju,Liu:2024fly,Arteaga:2024vde,Goncalves:2024lrk,Baratella:2025dum,Cao:2025jwb,Kierkla:2025vwp,Feng:2025wvc,Kierkla:2026bnm,Feng:2026iqo}.
As confirmed in the context of the $U(1)_{B-L}$ model in Ref.~\cite{Iso:2017uuu}, there is a possibility that the $U(1)_{X}$ phase transition can proceed after the QCD phase transition. 
Since the EW symmetry is not broken before the $U(1)_{X}$ symmetry breaking, it implies that all quarks do not acquire their current masses. 
In such a case, the QCD phase transition is expected to be first-order~\cite{Pisarski:1983ms}. 
Based on these findings, the dynamics of chiral symmetry breaking with multiple massless quarks has been numerically analyzed based on lattice simulations~\cite{Brown:1990ev, Cuteri:2021ikv}. 
Furthermore, in Ref.~\cite{Schmitt:2024pby}, it has been shown that a new type of phase transition called the tachyonic phase transition may occur if the $U(1)_{B-L}$ and the EW symmetry breaking are not induced even if the QCD phase transition occurs. 

During the tachyonic phase transition, the inhomogeneous scalar field excitation can form, and its dynamics may leave a characteristic stochastic gravitational wave (GW) signal~\cite{Dufaux:2007pt, Dufaux:2008dn}. 
This GW source is independent from that produced by the dynamics of vacuum bubble walls nucleated by cosmological first-order phase transitions \cite{Caprini:2024hue}. 
This possibility opens a novel path to explore new physics with the classical scale invariance. 

In this paper, we investigate the phase transition dynamics in the $U(1)_{X}$ extension of the SM and show that the tachyonic phase transition can be realized. 
We show that observations of the resulting GW signal can explore the parameter region with small $U(1)_{X}$ couplings where collider experiments and observations of GWs from first-order phase transitions cannot explore. In this model we have three RHNs and their Majorana masses are generated after the breaking of the $U(1)_X$ symmetry via CW mechanism. Following the electroweak symmetry breaking the light neutrino mass is generated by the seesaw mechanism further inducing the mixing between the light and heavy neutrino mass eigenstates. In this work we consider a situation where three RHNs are degenerate in nature with two benchmark masses of $m_{N}=1$ TeV and 10 TeV. After the $U(1)_X$ symmetry breaking, the mass of new neutral gauge boson $Z^\prime$ is induced. In particular, if the mass of the new gauge boson $Z'$ and the $U(1)_{X}$ gauge coupling are within the range $10^{6}\,{\rm GeV} < m_{Z'} < 10^{7}\,{\rm GeV}$ and $0.001 < g_{X} < 0.25$, the GW signal from the tachyonic phase transition may be tested by DECIGO~\cite{Kawamura:2006up,Yagi:2011wg}.
Our results further emphasize the importance of the complementarity between colliders and GW observations.

This paper is organized as follows. In Sec.~\ref{secII} we describe the model. The effective potential at finite temperature has been discussed in Sec.~\ref{secIII}. We analyzed $U(1)_X$ phase transition in Sec.~\ref{secIV} and GW from the breaking of $U(1)_X$ symmetry in Sec.~\ref{secV}, respectively. We discuss the results in Sec.~\ref{secVI} and finally conclude our paper in  Sec.~\ref{secVII}.
\section{General $U(1)_X$ extension under classical scale invariance}
\label{secII}
\begin{table}[t]
\begin{center}
\renewcommand{\arraystretch}{1}
\begin{tabular}{|c|c|c|c|c|}
\hline
 & $SU(3)_{C}$ & $SU(2)_{L}$ & $U(1)_{Y}$ & $U(1)_{X}$ \\
\hline
$q_L^i$ & $\mathbf{3}$ & $\mathbf{2}$ & $+1/6$
 & $x_q=\frac{1}{6}x_H+\frac{1}{3}x_\Phi$ \\
$u_R^i$ & $\mathbf{3}$ & $\mathbf{1}$ & $+2/3$
 & $x_u=\frac{2}{3}x_H+\frac{1}{3}x_\Phi$ \\
$d_R^i$ & $\mathbf{3}$ & $\mathbf{1}$ & $-1/3$
 & $x_d=-\frac{1}{3}x_H+\frac{1}{3}x_\Phi$ \\
\hline
$\ell_L^i$ & $\mathbf{1}$ & $\mathbf{2}$ & $-1/2$
 & $x_\ell=- \frac{1}{2} x_H - x_\Phi$ \\
$N_R^i$ & $\mathbf{1}$ & $\mathbf{1}$ & $0$
 & $x_\nu=- x_\Phi$ \\
$e_R^i$ & $\mathbf{1}$ & $\mathbf{1}$ & $-1$
 & $x_e = -x_H - x_\Phi$ \\
\hline
$H$ & $\mathbf{1}$ & $\mathbf{2}$ & $+1/2$ & $x_H/2$ \\
$\Phi$ & $\mathbf{1}$ & $\mathbf{1}$ & $0$ & $2x_\Phi$ \\
\hline
\end{tabular}
\end{center}
\caption{The particle content and the corresponding charge assignments in the $U(1)_{X}$ extension of the SM.}
\label{Tab:particle_contents}
\end{table}

We investigate a classically conformal and anomaly-free extension of the SM with the gauge group $SU(3)_C\otimes SU(2)_L\otimes U(1)_Y \otimes U(1)_X$. 
As summarized in Tab.~\ref{Tab:particle_contents}, the field content extends that of the SM by incorporating three generations of SM-singlet RHNs $(N_R^i)$ alongside a SM-singlet $U(1)_X$ scalar field $\Phi$.
In the following, we specify the covariant derivatives for the $U(1)_Y\otimes U(1)_X$ gauge sector as
\begin{equation}
D_\mu \equiv \partial_\mu  
			- i\left(\begin{array}{cc} Y_{1} & Y_{X} \end{array}\right )
			\left(\begin{array}{cc} g_{1} & g_{1X} \\g_{X1} & g_{X} \end{array}\right)
			\left(\begin{array}{c} B_{\mu} \\B_{\mu}^\prime \end{array}\right), 
 \label{Eq:covariant_derivative}
\end{equation}
where $Y_1$ and $Y_X$ are the $U(1)_Y$ and $U(1)_X$ charges of the corresponding
particle, and the off-diagonal couplings $g_{X1}$ and $g_{1X}$ are associated with the kinetic mixing between the $U(1)_Y$ and $U(1)_X$ gauge bosons.
Since an orthogonal rotation of $(B_\mu, B'_\mu)$ leaves the gauge kinetic terms
invariant, one of the off-diagonal couplings can be removed.
In the following, we work in the basis with $g_{X1}=0$ and define
\begin{align}
  g_Y \equiv g_1\,, \quad x_{\rm mix} \equiv g_{1X}/g_X\,,
\end{align}
Then, the covariant derivative is expressed by
\begin{align}
D_\mu 
= \partial_\mu - i g_Y Y_1 B_\mu - i g_X \left( Y_X + x_{\rm mix}\, Y_1 \right) B'_\mu \,.
\end{align}
Because $Y_X = x_H Y_1 + x_\Phi (B-L)$ for all the fields in Tab.~1, a nonzero
$x_{\rm mix}$ is equivalent to the shift $x_H \to x_H + x_{\rm mix}$ in the
couplings of $B'_\mu$.
Even if $x_{\rm mix}$ vanishes at a certain scale, it is in general induced by the RG evolution.

We calculate the generation-independent $U(1)_X$ charge assignments for the charged fermions of the model solving the gauge and gravitational anomaly-free conditions~\cite{Das:2016zue}
\begin{align}
\begin{aligned}
{\rm U}(1)^\prime \times \left[ {\rm SU}(3)_C \right]^2&\ :&
			2x_q - x_u - x_d &\ =\  0 \,,  \\
{\rm U}(1)^\prime \times \left[ {\rm SU}(2)_L \right]^2&\ :&
			3x_q + x_\ell &\ =\  0 \,, \\
{\rm U}(1)^\prime \times \left[ {\rm U}(1)_Y \right]^2&\ :&
			x_q - 8x_u - 2x_d + 3x_\ell - 6x_e &\ =\  0 \,, \\
\left[ {\rm U}(1)^\prime \right]^2 \times {\rm U}(1)_Y&\ :&
			x_q^2 - 2x_u^2 + x_d^2 - x_\ell^2 + x_e^2 &\ =\  0 \,,  \\
\left[ {\rm U}(1)^\prime \right]^3&\ :&
			6x_q^3 - 3x_u^3 - 3x_d^3 + 2x_\ell^3 - x_\nu^3 - x_e^3 &\ =\  0 \,, \\
{\rm U}(1)^\prime \times \left[ {\rm grav.} \right]^2&\ :&
			6x_q - 3x_u - 3x_d + 2x_\ell - x_\nu - x_e &\ =\  0 \,.
\label{Eq:anomaly_free_cond}
\end{aligned}
\end{align}

The observed fermion masses and flavor mixing can be reproduced from the following Yukawa interactions as
\begin{align}
\begin{aligned}
{\cal L}_{\rm Yukawa} = 
& - Y_u^{\alpha \beta} \overline{q_L^\alpha} \tilde{H} u_R^\beta
- Y_d^{\alpha \beta} \overline{q_L^\alpha} H d_R^\beta
- Y_e^{\alpha \beta} \overline{\ell_L^\alpha} H e_R^\beta \\
& - Y_\nu^{\alpha \beta} \overline{\ell_L^\alpha} \tilde{H} N_R^\beta
-\frac{1}{2}Y_N^\alpha \Phi \overline{(N_R^\alpha)^c} N_R^\alpha
+{\rm h.c.} \,, 
\end{aligned}
\label{LYk}
\end{align}
where $\tilde{H}=i\tau^2H^\ast$, with $\tau^2$ denoting the second Pauli matrix.
The fourth and fifth terms on the right-hand side of Eq.~\eqref{LYk} induce the seesaw mechanism after the spontaneous breaking of $U(1)_X$ and electroweak gauge symmetries through which tiny neutrino masses and flavor mixings are generated. Necessity of these Yukawa interactions further imposes the following conditions as
\begin{align}
\frac{1}{2}x_H = - x_q + x_u = x_q - x_d =  - x_\ell + x_\nu = x_\ell - x_e \,, \quad x_\Phi	= - x_\nu \,.
\label{cond1}
\end{align}
Solving Eqs.~\eqref{Eq:anomaly_free_cond} and \eqref{cond1}, we find that the $U(1)_X$ charges of the particles are determined by only two free parameters, $x_H$ and $x_\Phi$.
From the $U(1)_X$ charges of the particles, listed in Tab.~\ref{Tab:particle_contents}, the fact is manifested that the $U(1)_X$ gauge group can be defined as a linear combination of the SM $U(1)_Y$ and the $U(1)_{\rm B-L}$ gauge groups.
Since the $U(1)_X$ gauge coupling $g_X$ is a free parameter of the model and appearing as $x_H g_X$ or $x_\Phi g_X$, we fix $x_\Phi=1$ in our analysis without loss of generality which further excludes the case where $U(1)_X$ is identical to the SM $U(1)_Y$ gauge group.
Varying the parameter $x_H$ recovers several benchmark gauge frameworks as specific limits.
For instance, $x_H = -2$ yields the $U(1)_R$ model, where the left-handed quark and lepton doublets are uncharged, leaving the $Z'$ to couple exclusively to right-handed fermions.
Similarly, setting $x_H = -1$, $-1/2$, or $1$ neutralizes $e_R$, $u_R$, and $d_R$, respectively.
Since uncharged particles do not interact with the $Z'$ boson, these choices naturally generate chiral gauge interactions.
In contrast, $x_H = 0$ reproduces the vector-like $U(1)_{B-L}$ model, in which left- and right-handed fermions carry identical charges.
In this limit, the SM Higgs doublet is $U(1)_{\rm B-L}$ neutral, whereas the singlet scalar $\Phi$ carries a charge of 2. 
In this set-up, when we choose $x_H=-32/41$ the beta function of $g_{X1}$ ($g_{1X}$) at the 1-loop level only carries terms proportional to $g_{X1}$ ($g_{1X}$). The corresponding beta functions are given in Appendix \ref{Sec:RGEs}. This condition is called the orthogonal condition between the $U(1)_Y$ and $U(1)_X$ gauge groups at the 1-loop level. Under this condition $g_{X1}$ and $g_{1X}$ do not run from the initial condition $g_{X1}=g_{1X}=0$ set in this study at an energy scale. This is a good approximation even at the 2-loop level when $x_H$ is slightly larger than $-32/41$.
   
Imposing the classical scale invariance, the scalar potential for the SM Higgs $H$ and the $U(1)_{X}$ scalar field $\Phi$ can be given as~\cite{Das:2016zue}
\begin{align}
V_{0}(H, \Phi) = \lambda_H \left( H^\dagger H \right)^2 
	+ \lambda_\Phi \left( \Phi^\dagger \Phi \right)^2 
	+ \lambda_{\rm mix}  \left( H^\dagger H \right)  \left( \Phi^\dagger \Phi \right), 
\label{Eq:classical_potential}
\end{align}
where conformal invariance forbids the mass terms for the scalar fields. An approximation considering negligibly small mixing $(\lambda_{\rm mix})$ between the scalar fields allows us to analyze the scalar potential separately for $H$ and $\Phi$. 
When the Majorana Yukawa coupling $Y_N^\alpha$ are smaller than $g_X$, allowing $\Phi$ sector to be identical to the original Coleman-Weinberg \cite{Coleman:1973jx} scenario so that radiative breaking of the $U(1)_X$ symmetry could be fulfilled.
Once the $U(1)_{X}$ scalar field $\Phi$ acquires a vacuum expectation value (VEV), the tree-level mass term for the SM Higgs boson is generated through the mixing between $H$ and $\Phi$ as given in Eq.~(\ref{Eq:classical_potential}).
We consider the case with $\lambda_{\rm mix} <0$ so that the mass-squared of the SM Higgs doublet becomes negative after the $U(1)_{X}$ symmetry breaking. 
As a result, the electroweak symmetry breaking is also simultaneously induced.

We first analyze the $U(1)_X$ scalar sector assuming negligibly small mixing between $H$ and $\Phi$ fields. In the absence of a mass term, the Coleman-Weinberg potential at 1-loop level can be written as~\cite{Coleman:1973jx} 
\begin{align}
V(\varphi) =  \frac{\lambda_\Phi}{4} \varphi^4 
+ \frac{\beta_\Phi}{8} \varphi^4 \left[  \ln \left( \frac{\varphi^2}{v_\varphi^2} \right) - \frac{25}{6} \right] \,,
\label{Eq:CW_potential} 
\end{align}
where $\varphi/\sqrt{2} = \Re[\Phi]$, and $v_\varphi$ denotes the VEV of the $U(1)_{X}$ scalar field. 
We here choose the renormalization scale as $\mu = v_{\varphi}$. 
The coefficient of the 1-loop quantum corrections $\beta_{\Phi}$ is given by 
\begin{align}
\beta_\Phi
&= \frac{1}{16 \pi^2} \left[ 20\lambda_\Phi^2 + 6 \left ( 4 x_\Phi^2 g_{X1}^2 + 4 x_\Phi^2 g_{X}^2 \right)^2 - \sum_i(Y_N^i)^4 \right] \\
& \simeq  \frac{1}{16 \pi^2} \left[ 6 \left(2 x_\Phi g_{X} \right)^4 - \sum_i(Y_N^i)^4 \right], ~~~({\rm considering}~ \lambda_\Phi^2 \ll (2 x_\Phi g_{X})^4) 
\end{align}
where we took $g_{X1}=g_{1X}= 0$ at $ \langle \varphi \rangle = v_\varphi$ for simplicity. 
The stationary condition $ dV (\varphi)/d\varphi |_{\varphi=v_\varphi} = 0$ from Eq.~(\ref{Eq:CW_potential}) yields
\begin{eqnarray}
   \lambda_\Phi = \frac{11}{6} \beta_\Phi, 
\label{eq:stationary}
\end{eqnarray} 
where $\lambda_\Phi$ denotes the renormalized self-coupling of $\varphi$. 
According to Ref.~\cite{Khoze:2014xha} it can be defined as 
\begin{eqnarray}
 \lambda_\Phi = \frac{1}{3 !}\left. \frac{d^4V(\varphi)}{d \varphi^4} \right|_{\varphi=v_\varphi}. 
\end{eqnarray}  
Associated with the radiative $U(1)_X$ symmetry breaking and followed by the breaking of electroweak symmetry, a new gauge boson $Z^\prime$ obtains the following mass
\begin{align}
  m_{Z^\prime}  = \sqrt{(2 x_\Phi g_{X} v_\varphi)^2  +  \left( \frac{x_H}{2} g_{X} v \right)^2} \simeq 2 g_{X} v_\varphi\,,
\end{align} 
where we assumed $v_\varphi \gg v$ with $v=246\,{\rm GeV}$. 

On the other hand, the $U(1)_X$ scalar field mass is given by 
\begin{eqnarray}
m_\varphi^2 
= \left. \frac{d^2 V}{d\varphi^2}\right|_{\varphi=v_\varphi}  
\simeq \frac{3}{8 \pi^2}  \frac{ m_{Z^\prime}^4 - 2 m_N^4}{v_\varphi^2} \,.
\label{Eq:mass_phi}
\end{eqnarray} 
For negligibly small $Y_N$, Eq.~(\ref{Eq:mass_phi}) reduces to well-known Coleman-Weinberg relation~\cite{Coleman:1973jx}.
The potential minimum disappears when $m_N >  m_{Z^\prime}/2^{1/4}$ providing an upper bound on the heavy neutrino mass to realize the radiative $U(1)_X$ symmetry breaking.
To avoid destabilization of the $U(1)_X$ potential, we mainly consider the case with $m_{Z^\prime} \gg m_N$.
This does not imply that the Majorana type heavy neutrinos are very light.
Destabilization of the Higgs potential can be avoided even if there is a factor difference between $Z^\prime$ and heavy neutrino masses. 
At the renormalization scale $\mu=v_\Phi$, the contribution of $Y_N$ to the RG running is negligible~\cite{Iso:2009ss,Iso:2009nw,Das:2015nwk, Oda:2015gna,Das:2016zue}. 

Now we consider the electroweak symmetry breaking dynamics.  
After the $U(1)_X$ symmetry breaking, the negative quadratic term for the SM Higgs field can be produced through $\lambda_{\rm mix}$ in Eq.~(\ref{Eq:classical_potential}) as~\cite{Iso:2009ss,Iso:2009nw,Das:2015nwk, Oda:2015gna,Das:2016zue}
\begin{equation}
  V(h) = \frac{\lambda_H}{4}h^4 + \frac{\lambda_{\rm mix}}{4} v_\varphi^2 h^2 \,, 
\end{equation}
where $H = 1/\sqrt{2}\, (0\,,h)^{\rm T}$. Using the stationary condition $dV(h)/dh |_{h= v} =0$ at the tree level, we obtain 
\begin{align}
\lambda_{\rm mix}= - 2 \lambda_H \frac{v^2}{v_{\varphi}^2} \,. 
\end{align}
Hence, the Higgs mass can be calculated as 
\begin{align}
m_h^2= \left. \frac{d^2 V(h)}{dh^2}\right|_{h= v} = - \lambda_{\rm mix} v_{\varphi}^2 = 2 \lambda_H v^2.
\end{align}
which has been used as a boundary condition for $\lambda_{\rm mix}$ in the RGE analysis~\cite{Iso:2009ss,Iso:2009nw,Das:2015nwk, Oda:2015gna,Das:2016zue} fixing the renormalization scale at $\mu=v_{\varphi}$. 
If we consider the case with $\lambda_H\approx \mathcal{O}(0.1)$ and $v_{\varphi} \approx \mathcal{O}(10)\,{\rm TeV}$ satisfying the current collider constraints yields $|\lambda_{\rm mix}| \lesssim 10^{-5}$, which is extremely small~\cite{Electroweak:2003ram,ALEPH:2013dgf}.
Due to the current searches on heavy resonance~\cite{ATLAS:2019erb,CMS:2019tbu,CMS:2024bni,ATLAS:2024fdw}, we find that $m_{Z^\prime}\leq 6$ TeV is tightly constrained from the non-observing dilepton and dijet searches. As a result, the $\beta_\Phi$ term will dominate in $\lambda_\Phi$ compared to the contribution of small $\lambda_{\rm mix}$.

Under the $U(1)_X$ scenario the interaction between the $Z^\prime$ boson and the SM fermions $(f_{L(R)})$ can be written as 
\begin{align}
\mathcal{L} = -g_X (\overline{f_L}\gamma^\mu q_{f_{L}^{}}^{}  f_L+ \overline{f_R}\gamma^\mu q_{f_{R}^{}}^{}  f_R) Z_\mu^\prime \,,
\label{Lag1}
\end{align}
where $q_{f_{L(R)}^{}}^{}$ is the $U(1)_X$ charge of the corresponding left- (right-) handed fermion. 
Hence we calculate the partial decay widths of $Z^\prime$ into the SM charged fermions as
\begin{align}
\label{eq:width-ll}
\Gamma(Z^\prime \to \bar{f} f)
= N_C^{} \frac{m_{Z^\prime}^{} g_{X}^2}{24 \pi} \left[ \left( q_{f_L^{}}^2 + q_{f_R^{}}^2 \right) \left( 1 - \frac{m_f^2}{M_{Z^\prime}^2} \right) + 6 q_{f_L^{}}^{} q_{f_R^{}}^{} \frac{m_f^2}{m_{Z^\prime}^2} \right]
\sqrt{1-\frac{4 m_f^2}{m_{Z^\prime}^2}} \,, 
\end{align}    
where $m_f$ is the mass of the SM fermions and $N_C^{}=1\,(3)$ is the color factor for SM leptons (quarks). 
Neglecting the tiny neutrino mass, the partial decay width of $Z^\prime$ into a pair of single generation light neutrino is given by
\begin{align}   
\label{eq:width-nunu}
\Gamma(Z^\prime \to \nu \nu)
= \frac{m_{Z^\prime}^{} g_{X}^2}{24 \pi} q_{f_L^{}}^2~ \,,
\end{align} 
where $q_{f_L^{}}^{}$ is the $U(1)_X$ charge of the SM lepton doublet. 

After the radiative breaking of the $U(1)_X$ symmetry, the Majorana masses of the RHNs are generated as
\begin{align}
m_{N^\alpha} = Y_N^\alpha v_\varphi/\sqrt{2} \,. 
\label{Nmass}
\end{align}
The Dirac masses of the neutrinos are generated as $m_D^{\alpha\beta}=Y_\nu^{\alpha\beta}v/\sqrt{2}$ after electroweak symmetry breaking. 
Hence the resulting neutrino mass matrix can be written as
\begin{align}
m_\nu=
\begin{pmatrix}
0 & m_D \\
m_D^T & m_N
\end{pmatrix} \,. 
\label{num-1-1} 
\end{align}
Within the seesaw limit $(m_D\ll m_N)$, the light neutrino mass can be yielded as
\begin{align}
m_\nu\simeq -m_D m_N^{-1}m_D^T \,.
\label{nut}
\end{align}
As a result, the light and heavy neutrino mass eigenstates are mixed and the mixing angle is defined as  $m_D/m_N$. Through the small admixture with the SM light neutrinos, the heavy neutrinos interact with the SM gauge bosons. In this minimal scenario, these three generations of heavy neutrinos participate in the generation of tiny neutrino mass and observed flavor mixing. Hence we express the light neutrino $(\nu)$ flavor eigenstate in terms of the light $(\nu_m)$ and heavy $(N_m)$ neutrino mass eigenstates following 
\begin{align}
\nu \simeq  \nu_m  + V_{\ell N} N_m \,, 
\end{align}
where $V_{\ell N}=m_D/m_N$ being light-heavy neutrino mixing assuming $|V_{\ell N}| \ll 1$. Implementing $\nu$ in terms of the mass eigenstates, the charged current (CC) interaction can be written as
\begin{align}
\mathcal{L}_{\rm CC} \supset -\frac{g}{\sqrt{2}} W_{\mu}
\bar{\ell} \gamma^{\mu} P_L   V_{\ell N} N_m  + {\rm H.c.}, 
\label{CC}
\end{align}
where $\ell$ represents the three generations of the charged leptons from the SM in the vector form, and we write $P_L =\frac{1}{2} (1- \gamma_5)$ being the projection operator. Then the neutral current (NC) interaction in terms of the mass eigenstates can be written as 
\begin{align}
\mathcal{L}_{\rm NC} \supset 
-\frac{g}{2 \cos\theta_{\rm W}}  Z_{\mu} 
\left[ 
\overline{N_m} \gamma^{\mu} P_L  |V_{\ell N}|^2 N_m 
+ \left\{ 
\overline{\nu_m} \gamma^{\mu} P_L V_{\ell N}  N_m 
+ {\rm H.c.} \right\} 
\right] , 
\label{NC}
\end{align}
where $\theta_{\rm W}$ is the weak mixing angle. 
In the $U(1)_X$ extension of the SM, the interaction between the $Z^\prime$ and heavy neutrinos can be written by
\begin{align}
\mathcal{L}_N= -\frac{1}{2}g_X q_{N_R} \overline{N} \gamma^\mu \gamma_5 N Z_{\mu}^\prime.
\label{neut}
\end{align}
where $q_{N_R}$ is the $U(1)_X$ charge of heavy neutrinos. Hence we obtain the  partial decay width of $Z^\prime$ into a pair of single generation heavy neutrino as
\begin{align}
\label{eq:width-NN}
\Gamma(Z^\prime \to N^\alpha N^\alpha)
= \frac{ g_{X}^2 m_{Z^\prime}^{}}  {24 \pi} q_{N_R^{}}^2 \left( 1 - 4\frac{ m_{N^\alpha}^2}{m_{Z^\prime}^2} \right)^{\frac{3}{2}} \,.
\end{align}
In this paper we assume degenerate masses of the heavy Majorana neutrinos for simplicity where $Y_N^\alpha = Y_N$ which is equivalent to $m_{N^\alpha}= m_N$ for $\alpha=1,2,3$.

To compare our result with existing collider bounds, we first evaluate the LEP constraints on our model parameters for different $x_H$ considering $M_{Z^\prime} \gg \sqrt{s}$. Following the effective operator approach discussed in Refs.~\cite{Eichten:1983hw, Electroweak:2003ram,ALEPH:2013dgf}, we parametrize the pair production process $e^+e^-\to f\bar{f}$
by contact interactions as
\begin{align}
{\cal L}_{\rm eff} \ = \ \frac{g_X^2}{(1+\delta_{ef}) (\Lambda_{AB}^{f\pm})^2} \sum_{A,B=L,R}\eta_{AB}(\overline{e} \gamma^\mu P_A e)(\overline{f} \gamma_\mu P_B f) \,,
\label{eq1}
\end{align}
where 
$\delta_{ef}=1\,(0)$ for $f=e$ ($f\neq e$),  $\eta_{AB}=\pm 1$ or 0, and $\Lambda_{AB}^{f\pm}$ being the scale of the contact interaction, having constructive(destructive) interference by plus (minus) sign with the $e^+e^-\to f\bar{f}$ process within the SM~\cite{Kroha:1991mn,Carena:2004xs}. 
We calculate the matrix element for the $e^+e^-\to f\bar{f}$ process through the $Z^\prime$ exchange as
\begin{align}
\frac{g_X^2}{{m_{Z^\prime}}^2-s} [\overline{e} \gamma^\mu ({x_\ell}^\prime P_L+ {x_e}^{\prime} P_R) e] [\overline{f} \gamma_\mu (x_{f_L} P_L+x_{f_R} P_R) f] \,,
\label{eq2}
\end{align}
where ${x_{\ell(e)}}^\prime$ and $x_{f_{L(R)}}$ are the $U(1)_X$ charges of $e_{L(R)}$ and $f_{L(R)}$ respectively following Tab.~\ref{Tab:particle_contents}. 
Considering $m_{Z^\prime}^2 \gg s$ where $\sqrt{s}=209$ GeV for LEP-II and matching Eqs.~(\ref{eq1}) and (\ref{eq2}), we obtain a relation 
\begin{align}
m_{Z^\prime}^2  \ \gtrsim \ \frac{{g_X}^2}{4\pi} |{x_{e_A}} x_{f_B}| (\Lambda_{AB}^{f\pm})^2 \,,
\label{Lim}
\end{align}
which serves as a constraint condition for $m_{Z^\prime}/ g_X$ at the 95\% confidence level (CL) for different $U(1)_X$ charges taking $\Lambda_{AB}^{f\pm}$ reported in Ref.~\cite{ALEPH:2013dgf} using hadronic and leptonic final states where $AB=$LL, RR, LR, RL, VV (vector) and AA (axial-vector) interactions for $x_H=-2,-1,0,2$. 
These bounds are used in our numerical calculations. 
From our estimation, we find that LEP provides stringent constraints on $g_X$ for $m_{Z^\prime} \geq 6$ TeV. 

We calculate limits on $m_{Z^\prime}-g_X$ plane for $x_H=-2,-1,0,2$ using the dilepton \cite{ATLAS:2019erb,CMS:2019tbu} and dijet \cite{CMS:2018mgb,ATLAS:2019bov} 
final states from $Z^\prime$ production at the LHC at $\sqrt{s}=13$ TeV with $139$ fb$^{-1}$ integrated luminosity. We calculate the $Z^\prime$-mediated dilepton and dijet cross sections in our model $(\sigma_{\rm{Model}})$ for a trial $U(1)_X$ coupling $g_{\rm Model}$ estimating branching ratio of $Z^\prime$ into dilepton and dijet channels and compare with observed corresponding cross sections from LHC $(\sigma_{\rm Obs})$ at the 95\% C.L. to derive an upper bound on $U(1)_X$ coupling  
\begin{align}
g_X \ = \ \sqrt{g_{\rm{Model}}'^2 \left(\frac{\sigma_{\rm{Obs.}}}{\sigma_{\rm{Model}}}\right)} \,,
\label{gp}
\end{align}
We find that dilepton bounds are stronger than dijet bounds obtained from LHC. LHC provides the strongest bound on $g_X$ within $150$ GeV $\leq m_{Z^\prime} \leq 6$ TeV.

\section{Effective potential at finite temperatures}
\label{secIII}

Discussing the dynamics of symmetry breaking, if the $U(1)_{X}$ symmetry scale is sufficiently larger than the EW scale, we can treat these symmetry breaking dynamics independently~\cite{Chiang:2017zbz}. 
Since the same treatment is applicable to our scenario, we use this simplification in the following analysis. 

In the $U(1)_{X}$ model with the classical scale invariance, the finite temperature potential including the renormalization group (RG) improvement upto one-loop level is given by
\begin{align}
\label{eq:Veff_total}
V_{\rm eff}(\varphi) = V_{0}(\varphi) + V_{T}(\varphi, T) \,. 
\end{align}
The first term $V_{0}(\varphi)$ represents the RG-improved potential at the zero temperature, which is defined by 
\begin{align}
V_{0}(\varphi) = \frac{\lambda_{\Phi}(t)}{4} \varphi^4 \,,
\end{align}
with $t = \ln (\mu/v_{\varphi})$, where $\mu$ denotes the RG scale. 
The RG equations we solve are summarized in Appendix~\ref{Sec:RGEs}. 
For the value of $\mu$, we focus on the following range as taken in Refs.~\cite{Iso:2009ss, Schmitt:2024pby}
\begin{align}
\mu = \max \left( \varphi, \Lambda_{\rm QCD} \right) \,, 
\end{align}
where $\Lambda_{\rm QCD} = 0.1\,{\rm GeV}$.

The second term in Eq.~\eqref{eq:Veff_total}, which represents the thermal contribution at the one-loop level, is given by~\cite{Dolan:1973qd}
\begin{align}
\label{eq:VT}
V_{T}(\varphi, T) = \frac{T^4}{2\pi^2} \sum_{j} n_{j} J_{T}( M_{j}^2(\varphi, T)) \,, 
\end{align}
with
\begin{align}
    J_{T}(a_i^2)&=\int_0^{\infty} dx x^2 \ln \left[1 \mp {\rm exp}\left(-\sqrt{x^2 +a_i^2/T^2}\right)\right] \,, 
\end{align}
where the negative (positive) sign is for bosons (fermions). 
For the summation in Eq.~\eqref{eq:VT}, it runs over the $U(1)_{X}$ gauge boson $B'$, $U(1)_{Y}$ gauge boson $B$, three RHNs, the $U(1)_{X}$ scalar boson and the Nambu-Goldstone bosons. 
The factor $n_{i}$ denotes the degrees of freedom for the particle $i$.
The thermal masses for each field component are given by 
\begin{align}
\begin{aligned}
&M_{B'_{T}}^2(\varphi, T) = 4 g_{X}^2(t) \varphi^2 \,, \quad
M_{N_{i}}^2(\varphi, T) = \frac{(Y_{N}^{i}(t))^2}{2} \varphi^2 \,, \quad \\
&M_{\varphi}^2(\varphi, T) = 3 \lambda_{\Phi}(t) \varphi^2 + \Pi_{\varphi}(T) \,, \quad 
M_{\rm NG}^2(\varphi, T) = \lambda_{\Phi}(t) \varphi^2 + \Pi_{\varphi}(T) \,.
\end{aligned}
\end{align}
with 
\begin{align}
\label{eq:Pi_varphi}
&\Pi_{\varphi}(T) =
T^2 \left[ \frac{\lambda_{\Phi}(t)}{3} + \frac{\lambda_{\rm mix}(t)}{6} + g_{X}^2(t) + \frac{1}{24} {\rm Tr} \left[ Y_{N}^{\dagger}(t) Y_{N}(t) \right] \right] \,,
\end{align}
where $B'_{T}$ denotes the transverse mode of the $U(1)_{X}$ gauge field. 
We evaluate the effective potential in Landau gauge.

For the longitudinal modes of the $U(1)_{X}$ gauge boson $B'_{L}$ and $U(1)_{Y}$ gauge bosons $B_{L}$, there is a mass mixing effect due to the thermal correction. 
The corresponding mass matrix is given by 
\begin{align}
M_{B'_{L}B_{L}}^2 (\varphi, T) = 
\begin{pmatrix}
4 g_{X}^2(t) \varphi^2 + \Pi_{B'B'}(T) & \Pi_{B'B}(T) \\ 
\Pi_{BB'}(T) & \Pi_{BB}(T) 
\end{pmatrix} \,, 
\end{align}
with 
\begin{align}
\label{eq:Pi_BpBp}
&\Pi_{B'B'}(T) = g_{X}^2(t) T^2 
\left( \frac{11}{6} x_{H}^2 + \frac{11}{3} x_{H} x_{\rm mix}(t) + \frac{11}{6} x_{\rm mix}^2(t) + \frac{8}{3} x_{H} + \frac{8}{3} x_{\rm mix}(t) + 4  \right) \,, \\ 
\label{eq:Pi_BBp}
&\Pi_{BB'}(T) = \Pi_{B'B}(T) = g_{X}(t) g_{Y}(t) T^2 
\left( \frac{11}{6} x_{H}  + \frac{11}{6} x_{\rm mix}(t) + \frac{4}{3} \right) \,, \\
&\Pi_{BB}(T) = \frac{11}{6} g_{Y}^2(t) T^2 \,.
\end{align}
Therefore, the mass eigenvalues for longitudinal components of the $U(1)$ gauge fields $(X_{L}^{1}, X_{L^{2}})$ are given by 
\begin{align}
M_{X_{L}^{i}}^2 (\varphi, T) = \frac{1}{2} \left( M_{X}^2(\varphi, T) + \Pi_{BB}(T) \pm \sqrt{ \left\{ M_{X}^2(\varphi, T) - \Pi_{BB}(T) \right\}^2 + 4 \Pi_{BB'}^2(T) } \right) \,, 
\end{align}
with $M_{X}^2(\varphi, T) = 4 g_{X}^2(t) \varphi^2 + \Pi_{B'B'}(T)$. 
In our analysis, we use the mass eigenstates $(X_{L}^{1}, X_{L^{2}})$ for the longitudinal mode of the $U(1)$ gauge fields.

\section{Analysis for the $U(1)_{X}$ phase transition}
\label{secIV}

We explain the bubble nucleation dynamics using the effective potential discussed in the previous section. At high temperatures, the $U(1)_{X}$ symmetry is restored due to thermal symmetry restoration~\cite{Kirzhnits:1972ut, Dolan:1973qd, Weinberg:1974hy}. 
As the universe cools down, the $U(1)_{X}$ symmetry is broken at a certain temperature. 
When the phase transition is first-order, its dynamics is determined by the bubble nucleation rate~\cite{Linde:1981zj}
\begin{align}
\Gamma(T) \simeq T^4 \left[ \frac{S_{3}(T)}{2\pi T} \right]^{3/2} e^{-S_{3}(T)/T} \,, 
\end{align}
where $S_{3}(T)$ is the 3-dimensional action for bounce solutions $\phi$, which is defined by 
\begin{align}
S_{3}(T) = 4 \pi \int dr r^2 \left[ \frac{1}{2} \left( \frac{d\phi}{dr} \right)^2 + V_{\rm eff}(\phi, T) \right] \,. 
\end{align}

The nucleation temperature $T_{n}$, at which a bubble is nucleated in a Hubble patch during the horizon time, is defined by 
\begin{align}
\Gamma(T_{n}) = H^4(T_{n}) \,,
\end{align}
where $H(T)$ is the Hubble parameter. 
The time (temperature) evolution of the Hubble parameter is determined by the Friedmann equation
\begin{align}
H^2(T) = \frac{\rho_{R}(T) + \rho_{V}(T)}{3M_{p\ell}^2}  \,, 
\end{align}
where $\rho_{R}(T)$ and $\rho_{V}(T)$ denote the energy components of the radiation and vacuum energy at a temperature $T$, respectively. 
$M_{p\ell}$ represents the reduced Planck mass. 

The nucleation temperature $T_{n}$ gives the typical temperature when the phase transition starts. 
However, in the conformal $U(1)_{X}$ model, the percolation temperature $T_{p}$ can be in principle drastically different from $T_{n}$ because of the supercooling phase transition. 
The percolation dynamics is determined by the fraction of false vacuum domains, which is given by~\cite{Ellis:2018mja}
\begin{align}
F(T) = \exp \left[ - \frac{4\pi}{3} \int_{T}^{T_{c}} \frac{dT'}{T'^4} \frac{\Gamma(T')}{H(T')} \left( \int_{T}^{T'} d\tilde{T} \frac{v_{w}}{H(\tilde{T})}  \right)^3 \right] \,, 
\end{align}
where $T_{c}$ represents the critical temperature. 
In our analysis, we define the percolation temperature by $F(T_{p}) = 0.7$.

In the conformal $U(1)_{X}$ model, the $U(1)_{X}$ symmetry breaking can occur after the QCD phase transition~\cite{Iso:2017uuu}. 
In such a case, the chiral symmetry breaking occurs around $T = T_{\rm QCD}$ and then it leaves an additional potential term due to the Yukawa interaction between the top quark and the SM Higgs field $h$: $y_{t} \braket{\bar{t} t} h$.
Therefore, after the chiral symmetry breaking, it gives the VEV $v_{\rm QCD}$ for the SM Higgs field. 
As a result, the scalar potential obtains an additional correction 
\begin{align}
V_{\rm eff}^{T<T_{\rm QCD}}(\varphi, T) = \frac{\lambda_{\rm mix}(t)}{4} v_{\rm QCD}^2 \varphi^2 + V_{\rm eff}(\varphi, T) \,, 
\end{align}
where the second term is given in Eq.~\eqref{eq:Veff_total}. 
In our numerical calculation, we take $v_{\rm QCD} = T_{\rm QCD} = 0.1\,{\rm GeV}$ as taken in Ref.~\cite{Marzo:2018nov}.

\section{Gravitational wave from the tachyonic $U(1)_{X}$ symmetry breaking}
\label{secV}

In several parameter regions in the conformally invariant $U(1)_{X}$ model, the percolation temperature can be lower than the QCD phase transition temperature $T_{\rm QCD}$. 
In such a case, there is a possibility that the $U(1)_{X}$ symmetry breaking proceeds via the classical rolling of the $U(1)_{X}$ scalar field instead of the bubble nucleation. 
This type of phase transition is called the tachyonic phase transition~\cite{Schmitt:2024pby}. 
The necessary condition for the tachyonic phase transition is given by
\begin{align}
\label{eq:Troll_Tp}
T_{\rm roll} > T_{p} \,, 
\end{align}
where $T_{\rm roll}$ is the classical rolling temperature, which is determined by 
\begin{align}
    \left. \frac{d^2 V_{\rm eff}(\varphi, T_{\rm roll})}{d \varphi^2} \right|_{\varphi = 0} = 0 \,. 
\end{align}
In our analysis, we regard $T_{\rm roll}$ as the temperature when the tachyonic phase transition occurs.

After the tachyonic phase transition occurs, the stored vacuum energy is released and the preheating process is realized. 
During this process, the $U(1)_{X}$ scalar field $\varphi$ passes through the tachyonic region with $\partial^2 V_{\rm eff}/\partial \varphi^2 <0$. 
This process is similar to the so-called tachyonic instability in cosmological inflation scenarios~\cite{Dufaux:2007pt, Dufaux:2008dn}. 
Considering the field fluctuation $\delta \varphi(t, {\bf x})$ for the $U(1)_{X}$ scalar field $\varphi(t, {\bf x})$ during this tachyonic phase transition, it is expressed by 
\begin{align}
\label{eq:varphi_phic_dphi}
\varphi(t, {\bf x}) = \varphi_{\rm cl}(t) + \delta \varphi(t, {\bf x}) \,, 
\end{align}
where $\varphi_{\rm cl}(t)$ represents the classical homogeneous component of the $U(1)_{X}$ scalar field. 
Performing the Fourier expansion for the field fluctuation $\delta \varphi(t, {\bf x})$, it can be expressed as 
\begin{align}
\label{eq:dphi_Fourier}
\delta \varphi(t, {\bf x}) =  \int \frac{d^3 \vec{k}}{(2\pi)^3}  \left[ \hat{a}_{k} u_{k}(t) e^{i \vec{k} \cdot {\bf x}} + {\rm h.c.} \right] \,,
\end{align}
where $\hat{a}_{k}\,(a^{\dagger}_{k})$ denotes the annihilation (creation) operator. 
The time dependence of $\delta \varphi(t, {\bf x})$ is described by the mode function $u_{k}(t)$ with a momentum $\vec{k}$.

On the other hand, the scalar field equation for the $\varphi$ field can be given in the Friedmann-Robertson-Walker spacetime by 
\begin{align}
\label{eq:EoM_varphi}
\ddot{\varphi} - \frac{1}{a^2(t)} \nabla^2 \varphi + 3 H(t) \dot{\varphi} + \frac{d V_{\rm eff}(\varphi, T)}{d \varphi} = 0 \,, 
\end{align}
where $H(t)$ and $a(t)$ denote the Hubble parameter and the scale factor, respectively. 
Substituting Eqs.~\eqref{eq:varphi_phic_dphi} and \eqref{eq:dphi_Fourier} into Eq.~\eqref{eq:EoM_varphi}, we obtain~\cite{Schmitt:2024pby}
\begin{align}
& \ddot \varphi_{\rm cl}(t) + 3 H(t) \dot \varphi_{\rm cl}(t) +  \frac{d V_{\rm eff}( \varphi = \varphi_{\rm cl}(t))}{d \varphi}  = 0 \,,  \\ 
& \ddot u_{k} + 3 H(t) \dot u_{k} + \omega_{k}^2 u_{k} = 0 \,, 
\end{align}
with 
\begin{align}
\label{eq:omega_k}
\omega_{k}^2 
= \frac{k^2}{a^2(t)} + \frac{d^2 V_{\rm eff}(\varphi = \varphi_{\rm cl}(t))}{d \varphi^2} 
\equiv \frac{k^2}{a^2(t)} + m_{\rm eff}^{2}(t)
\,, 
\end{align}
with 
\begin{align}
m_{\rm eff}^2(t) \equiv \frac{d^2 V_{\rm eff}(\varphi = \varphi_{\rm cl}(t))}{d \varphi^2} \,,
\end{align}
where we only took into account the linear order terms for the $\delta \varphi$ field because of $\delta \varphi \ll \varphi_{\rm cl}$. 

According to Eq.~\eqref{eq:omega_k}, the frequency $\omega_{k}$ can be imaginary numbers if the curvature of the potential can be a negative number. 
We note that such a situation is naturally realized during the tachyonic phase transition because the curvature around the origin of the scalar potential can be negative ($m_{\rm eff}^2|_{\varphi \simeq 0} < 0$). 
Therefore, during the tachyonic phase transition, the mode functions defined in the following range receive an exponential enhancement $u_{k} \propto e^{|\omega_{k}|t}$: 
\begin{align}
0 \leq \frac{k}{a} \leq \sqrt{ \left| m_{\rm eff}^2 (t)\right| }  \quad \text{with} \quad m_{\rm eff}^2(t) < 0 \,.
\end{align}
We denote this characteristic physical scale by
\begin{align}
R_* \sim \frac{a_*}{k_*}.
\end{align}
In many cases $k_*/a_*$ is of order the tachyonic mass scale, although the precise peak position depends on the time dependence of $m_{\rm eff}^2(t)$ and on the subsequent nonlinear dynamics.
In our analysis, we use the following formula as its validity has been confirmed in Ref.~\cite{Schmitt:2024pby}
\begin{align}
\frac{k_{*}}{a_{*}} \simeq \sqrt{ \left|\frac{\lambda_{\rm mix}}{2} v_{\rm QCD}^2 \right|} \equiv \Delta m_{\rm QCD} \,.
\end{align}

As discussed in Ref.~\cite{Schmitt:2024pby} in detail, in order to safely realize the classical scalar field rolling with evading the cosmological expansion effect, the following condition should be satisfied 
\begin{align}
\label{eq:mQCD_Hubble}
\Delta m_{\rm QCD} > H(T_{\rm roll}) \,.
\end{align}
We take into account this condition in the subsequent numerical analysis as a theoretical constraint for the tachyonic phase transition.

There are two thermal histories after the tachyonic phase transition.
During the tachyonic phase transition, the universe is vacuum energy dominant. 
Therefore, the Hubble parameter just after tachyonic amplification can be well approximated by 
\begin{align}
H_{*} 
\simeq \sqrt{ \frac{\rho_{V}}{3 M_{p \ell}^2}} 
\simeq \left( \frac{3 m_{Z'}^4 - 2 \sum_{i} m_{N_{i}}^4}{384\pi^2 M_{p \ell}^2} \right)^{1/2}  \,. 
\end{align}
If the total decay width of the $U(1)_{X}$ scalar field $\Gamma_{\varphi}$ satisfies $\Gamma_{\varphi} > H_{*}$, the inhomogeneous scalar field configuration can immediately decay after the tachyonic phase transition. 
However, in the case with $\Gamma_{\varphi} < H_{*}$, the scalar field excitation cannot decay and the universe becomes the matter dominant period. 
In this case, the reheating process occurs when the following condition is satisfied
\begin{align}
H_{\rm RH} = H_{*} \left( \frac{a_{*}}{a_{\rm RH}} \right)^{\frac32} = \Gamma_{\varphi} \,,
\end{align}
where $H_{\rm RH}$ and $a_{\rm RH}$ denote the Hubble parameter and the scale factor just after the reheating process, and $a_{*}$ is the scale factor after the tachyonic phase transition. 
As a result, the reheating temperature $T_{\rm RH}$ can be expressed as 
\begin{align}
\label{eq:definition_Trh}
\frac{\pi^2}{30} g_{\rm rh}^{*} T_{\rm RH}^4 = 3 M_{p \ell}^2 \min \left( H_{*}^2, \Gamma_{\varphi}^2 \right) \,. 
\end{align}
We use this formula as the definition of the reheating temperature, which is important when we discuss gravitational wave signals. 
In our model, the total decay width of the $U(1)_{X}$ scalar field is given by 
\begin{align}
\label{eq:Gamma_varphi}
\Gamma_{\varphi} = \sum_{i=1}^{3} \Gamma(\varphi \to N_{i}N_{i}) \theta(m_{\varphi} - 2m_{N_{i}})  + \Gamma(\varphi \to hh) \theta(m_{\varphi} - 2m_{h})
\end{align}
with
\begin{align}
\Gamma(\varphi \to N_{i}N_{i}) 
= \frac{|Y_{N_{i}}|^2}{32 \pi} m_{\varphi} \left( 1 - \frac{4m_{N_{i}}^2}{m_{\varphi}^2} \right)^{\frac32} \,, ~~
\Gamma(\varphi \to hh) 
&= \frac{\lambda_{\rm mix}^2 v_{\varphi}^2}{32 \pi m_{\varphi}} \sqrt{1 - \frac{4 m_{h}^2}{m_{\varphi}^2}} \,, 
\end{align}
where $\theta(x)$ represents the Heaviside function.

As shown in Eqs.~\eqref{eq:definition_Trh} and~\eqref{eq:Gamma_varphi}, the reheating process from the tachyonic phase transition is not effective when the following condition is not satisfied: 
\begin{align}
\label{eq:m_varphi_mh_mN}
m_{\varphi} > 2 m_{N_{i}} \quad \text{and} \quad m_{\varphi} > 2 m_{h} \,. 
\end{align}
In our analysis, we focus mainly on parameter regions satisfying Eq.~\eqref{eq:m_varphi_mh_mN}.

The connection to GWs follows from the tensor perturbation carried out by the inhomogeneous scalar field formed by the tachyonic phase transition. 
The spatial part of the scalar energy-momentum tensor is expressed by
\begin{align}
T_{ij}
= \partial_i\varphi\,\partial_j\varphi
+ \delta_{ij}
\left[
\frac{1}{2} \dot \varphi^2 - \frac{1}{2a^2}(\nabla \varphi)^2 - V_{\rm eff}(\varphi)
\right] \,. 
\end{align}
The trace and isotropic pressure terms do not contribute to the transverse-traceless source. 
Therefore, the relevant source for GWs is given by~\cite{Dufaux:2007pt}
\begin{align}
\label{eq:PiTT}
\Pi_{ij}^{\rm TT}(t, {\bf x})
=
\frac{1}{a^2}
\left[
\partial_i \delta \varphi(t, {\bf x})
\partial_j \delta \varphi(t, {\bf x})
\right]^{\rm TT}.
\end{align}

The tensor perturbations for the scalar field in Eq.~\eqref{eq:PiTT} induce the gravitational field fluctuation $h_{ij}$ which is determined by 
\begin{align}
h_{ij}''+2 H h_{ij}'-\nabla^2h_{ij}
=
\frac{2}{M_{p \ell}^2}\Pi_{ij}^{\rm TT} \,.
\end{align}
Since $\Pi_{ij}^{\rm TT}$ is quadratic in terms of $\delta \phi$ which is exponentially amplified during the tachyonic phase transition, the GW source is enhanced much more rapidly than the scalar mode functions themselves.

The subsequent fragmentation, collision, and relaxation of scalar field configurations generate a large, time-dependent anisotropic stress tensor~\cite{Dufaux:2007pt}.
Denoting the characteristic source size by $R_*\sim a_*/k_*$, we estimate the peak GW abundance at production as~\cite{Schmitt:2024pby}
\begin{align}
\Omega_{\rm peak,*}
\sim \xi_{\rm eff}(H_*R_*)^2 
\sim \xi_{\rm eff}\left(\frac{a_*H_*}{k_*}\right)^2 \,,
\end{align}
where $\xi_{\rm eff}$ encodes the efficiency of converting scalar gradient energy into gravitational waves.
Values of $\xi_{\rm eff}=0.05$ and $\xi_{\rm eff}\sim0.15$ have been obtained for hybrid inflation~\cite{Garcia-Bellido:2007nns} and chaotic inflation~\cite{Dufaux:2007pt}, respectively.
For cosmological first-order phase transitions, $\xi_{\rm eff}=\mathcal{O}(10^{-5}\text{--}10^{-1})$, depending on the GW sources, latent heat, and wall velocity~\cite{Caprini:2024hue}.
In our analysis, we adopt $\xi_{\rm eff}=10^{-2}$ as a representative value.

The peak frequency at production $f_{*}$ is set by the characteristic scale of the scalar inhomogeneities.
Taking the characteristic GW wavenumber to be of order $k_{*} \sim \Delta m_{\rm QCD}$ as discussed in Ref.~\cite{Schmitt:2024pby}, we estimate
\begin{align}
f_{*} \sim \frac{k_*}{a_*} \sim \Delta m_{\rm QCD} \,.
\end{align}
The present-day peak frequency is obtained by redshifting as
\begin{align}
f_0=\frac{a_*}{a_0}f_* \,.
\end{align}
Thus, the tachyonic mass scale sets the characteristic frequency at production, while the reheating history determines its subsequent redshift.

Following Ref.~\cite{Schmitt:2024pby}, the ratio of scale factors is given by
\begin{align}
r_{\rm RH}
\equiv
\frac{a_*}{a_{\rm RH}}
=
\min\left[1,\left(\frac{\Gamma_\varphi}{H_*}\right)^{2/3}\right] \,,
\end{align}
where we used the fact that the transient matter dominant universe is realized during the period between the GW production and the reheating process if $\Gamma_{\varphi} < H_{*}$. 
The present-day peak frequency can therefore be written as
\begin{align}
f_{\rm peak}^{\rm tach}
\simeq
1.65\times10^{-7}\,{\rm Hz}\,
\frac{f_*}{H_{\rm RH}}
\left(\frac{T_{\rm RH}}{{\rm GeV}}\right)
\left(\frac{g^*_{\rm rh}}{100}\right)^{1/6}
r_{\rm RH} \,,
\end{align}
where $H_{\rm RH}=\min(H_*,\Gamma_\varphi)$ and $g^*_{\rm rh}$ denotes the effective relativistic degrees of freedom at reheating.
Here, the effective degrees of freedom for energy and entropy are taken to be equal at reheating.
Including the subsequent redshift effect, the present-day peak amplitude is given by
\begin{align}
\Omega_{\rm peak}^{\rm tach}h^2
\simeq
1.67\times10^{-5}
\left(\frac{100}{g^*_{\rm rh}}\right)^{1/3}
r_{\rm RH}\,\Omega_{\rm peak,*} \,.
\end{align}

In our numerical analysis, we evaluate the quantities entering these expressions including the RHN contributions and determine the reheating temperature using Eq.~\eqref{eq:definition_Trh}. 
The peak frequency and amplitude are understood as phenomenological estimates; a precise determination of GW spectra requires nonlinear simulations.

\section{Results and discussions} \label{secVI}

In this section, we show our main numerical results obtained by using the framework summarized in the previous sections.

\begin{figure}[t]
    \centering
    \includegraphics[width=0.46\linewidth]{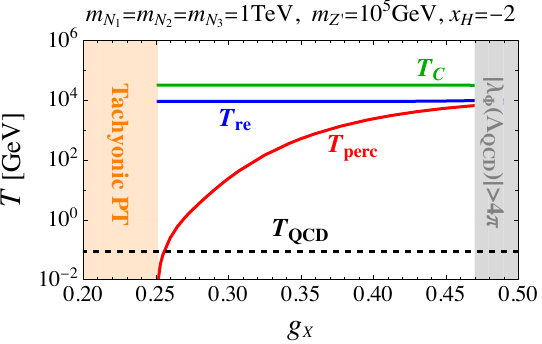}
    \includegraphics[width=0.46\linewidth]{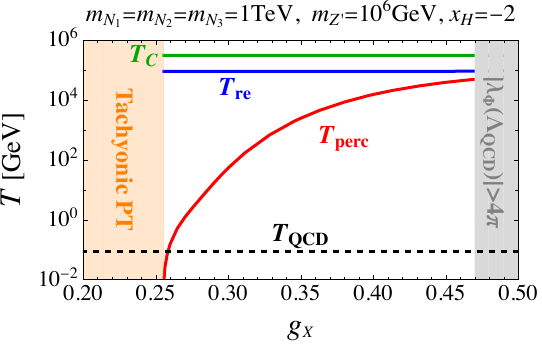}   
    \includegraphics[width=0.46\linewidth]{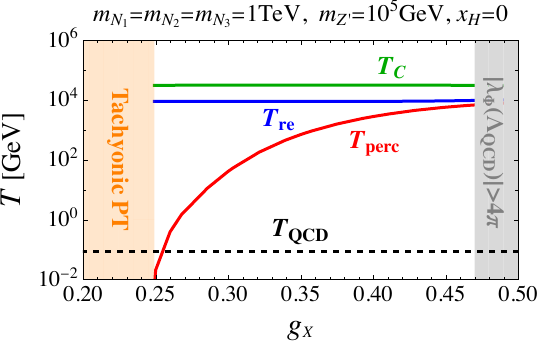}
    \includegraphics[width=0.46\linewidth]{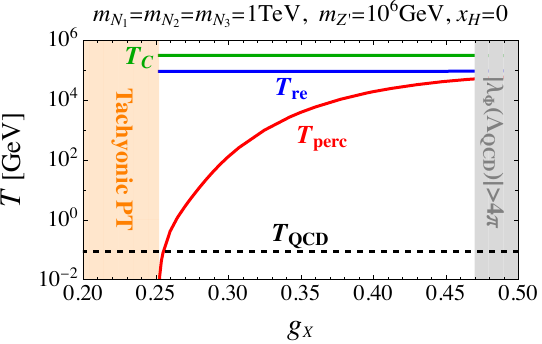}    
    \includegraphics[width=0.46\linewidth]{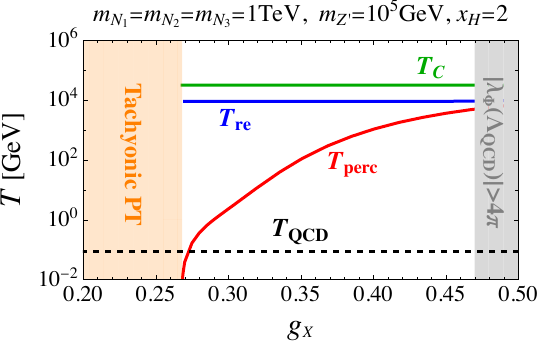}
    \includegraphics[width=0.46\linewidth]{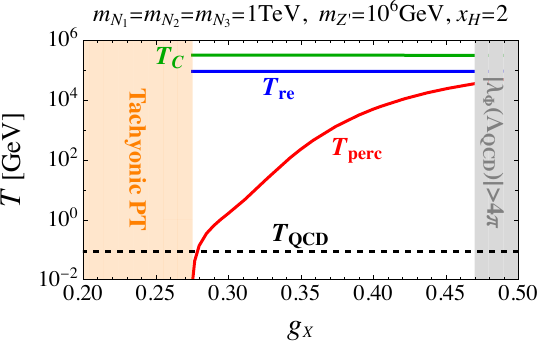}     
    \caption{
    The gauge coupling dependence of the characteristic temperatures relevant for the phase transition considering three degenerate heavy neutrinos of mass 1 TeV, two benchmark points of $m_{Z^\prime}=100$ (1000) TeV given in the left (right) panel for $x_H=-2$ (top, $U(1)_R)$, 0 (middle,B$-$L) and 2 (bottom). The red, blue and green solid lines denote the percolation $(T_{\rm perc})$, the reheating $(T_{\rm re})$ and the critical $(T_{C})$ temperatures, respectively. In the orange region, the tachyonic phase transition can occur. The perturbativity excludes the gray region. The black dashed line represents the temperature $(T_{\rm QCD})$ for the QCD phase transition.} 
    \label{fig:TgXplot-1}
\end{figure}
\begin{figure}[t]
    \centering
    \includegraphics[width=0.46\linewidth]{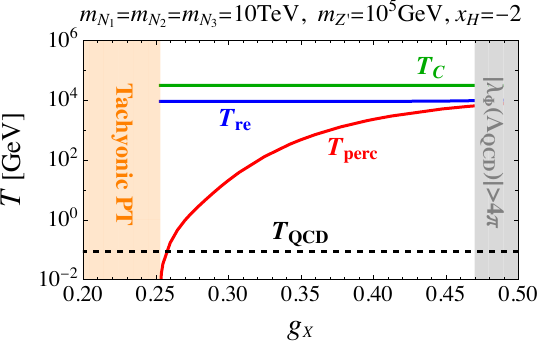}
    \includegraphics[width=0.46\linewidth]{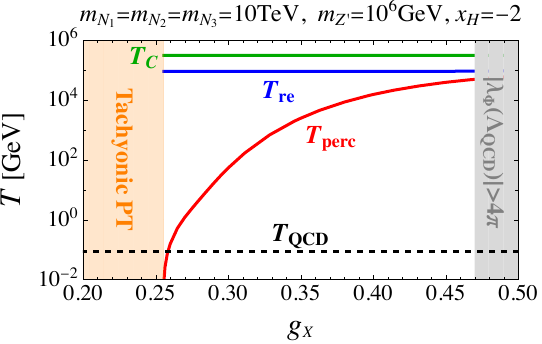}   
    \includegraphics[width=0.46\linewidth]{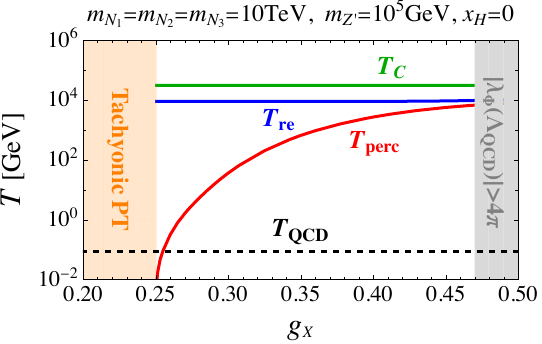}
    \includegraphics[width=0.46\linewidth]{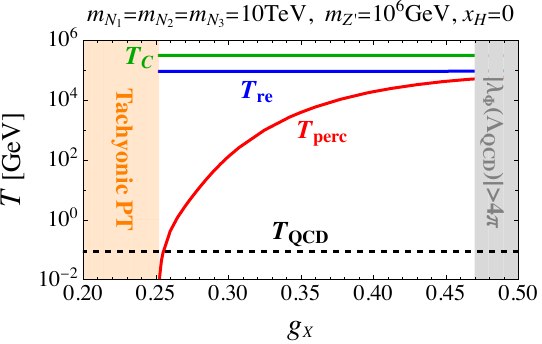}    
    \includegraphics[width=0.46\linewidth]{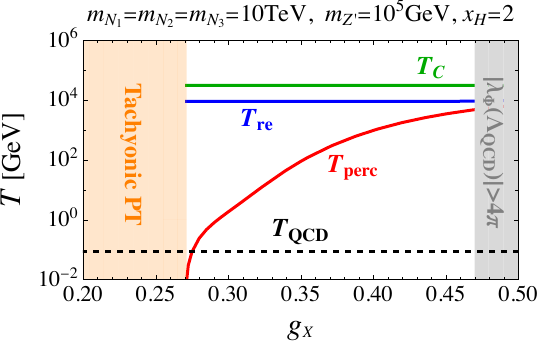}
    \includegraphics[width=0.46\linewidth]{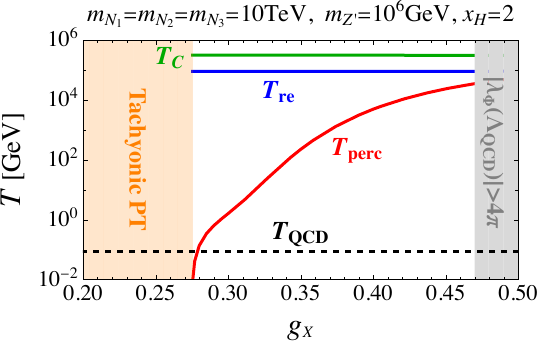}     
    \caption{
    The gauge coupling dependence of the characteristic temperatures relevant for the phase transition considering three degenerate heavy neutrinos of mass 10 TeV, two benchmark points of $m_{Z^\prime}=100$ (1000) TeV given in the left (right) panel for $x_H=-2$ (top, $U(1)_R$), 0 (middle, B$-$L) and 2 (bottom). The red, blue and green solid lines denote the percolation $(T_{\rm perc})$, the reheating $(T_{\rm re})$ and the critical $(T_{C})$ temperatures, respectively. In the orange region, the tachyonic phase transition can occur. The perturbativity excludes the gray region. The black dashed line represents the temperature $(T_{\rm QCD})$ for QCD phase transition.} 
    \label{fig:TgXplot-2}
\end{figure}

In Fig.~\ref{fig:TgXplot-1}, the $g_{X}$ dependence of several characteristic temperatures relevant for the $U(1)_{X}$ phase transition is shown. 
In the orange region, the requirement for the tachyonic phase transition given in Eq.~\eqref{eq:Troll_Tp} is satisfied. 
For the critical temperature $T_{c}$ (green), it is determined by the condition that the potential values at true and false vacua are equivalent at $T_{c}$. 
The reheating temperature is obtained by Eq.~\eqref{eq:definition_Trh}.
For the percolation temperature $T_{\rm perc}$, it is small if the value of $g_{X}$ is also small because the bounce action $S_{3}$ is proportional to $g_{X}^{-3}$ in the high temperature limit~\cite{Iso:2017uuu}. 
In particular, $T_{\rm perc}$ can be lower than the QCD phase transition temperature $T_{\rm QCD}$ if $g_{X} < 0.25$. 
It implies that the $U(1)_{X}$ and the EW symmetry breaking cannot be completed by the nucleation and expansion of vacuum bubbles formed by a first-order phase transition. 
Instead, in the range $g_{X} < 0.25$, the tachyonic phase transition plays an important role in completing the $U(1)_{X}$ phase transition in addition to the subsequent EW symmetry breaking. 
In addition, Fig.~\ref{fig:TgXplot-2} indicates that the $x_{H}$ dependence of the several phase transition temperature is modest because its contribution to the thermal potential only appears in the thermal correction to the field dependent masses for bosonic fields as shown in Eqs.~\eqref{eq:Pi_BpBp} and \eqref{eq:Pi_BBp}.

\begin{figure}[t]
\centering
\includegraphics[width=0.48\linewidth]{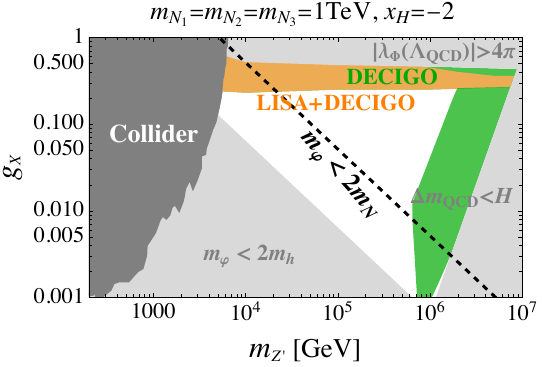}
\includegraphics[width=0.48\linewidth]{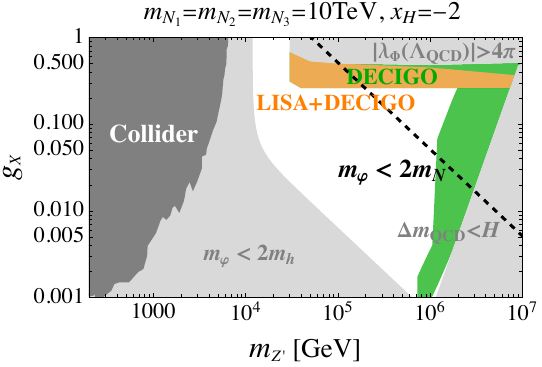} 
\includegraphics[width=0.48\linewidth]{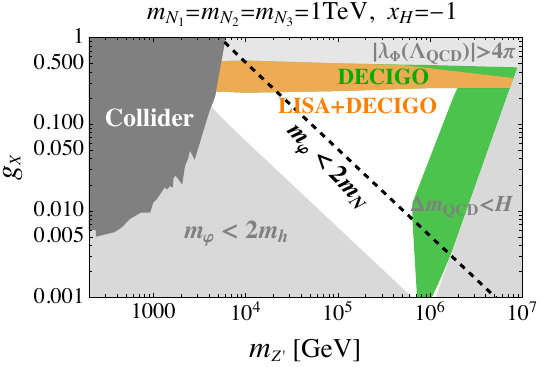}
\includegraphics[width=0.48\linewidth]{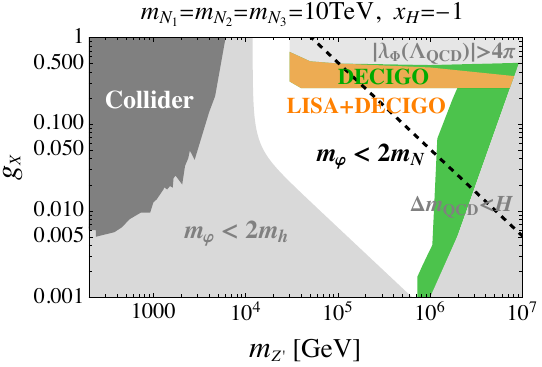} 
\caption{Parameter region explored by GW observatories and colliders for each benchmark point for $x_H=-2$ (top, $U(1)_R$), $-1$ (bottom) respectively for degenerate heavy neutrinos with masses 1 (10)TeV in the left (right) panel. In the colored regions, GWs can be detectable at each observatory. 
The dark gray region shows constraints by current collider searches from LEP and LHC as described in Sec.~\ref{secII}. 
In the light gray regions, the perturbativity bound $|\lambda_{\Phi}|<4\pi$ or the requirements from Eq.~\eqref{eq:m_varphi_mh_mN} and Eq.~\eqref{eq:mQCD_Hubble} are not satisfied.}
\label{fig:2dim_mZp_gX-1}
\end{figure}

\begin{figure}[t]
\centering
\includegraphics[width=0.48\linewidth]{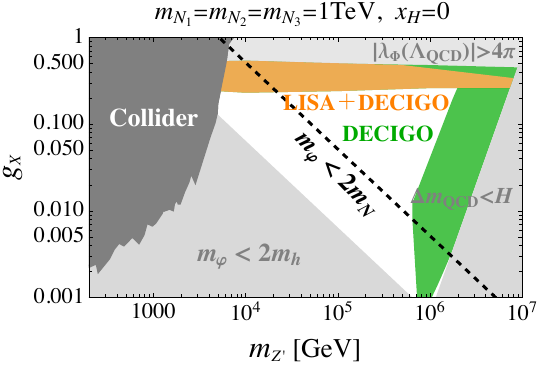}
\includegraphics[width=0.48\linewidth]{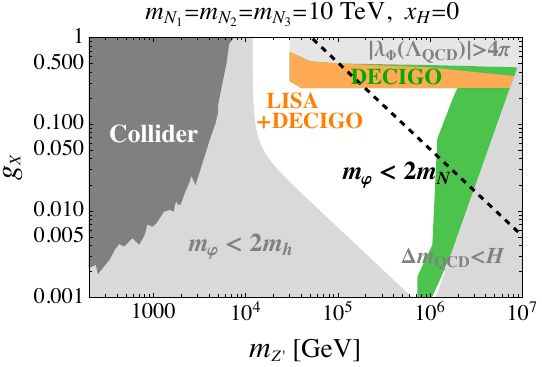}
\includegraphics[width=0.48\linewidth]{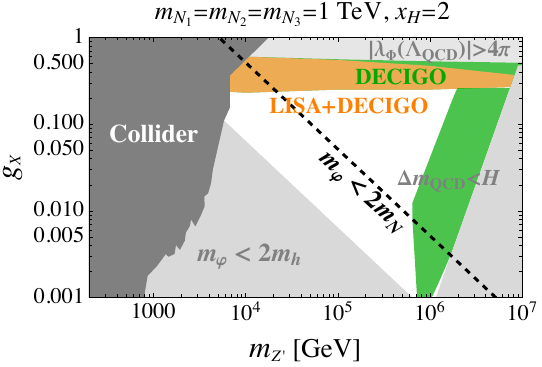}
\includegraphics[width=0.48\linewidth]{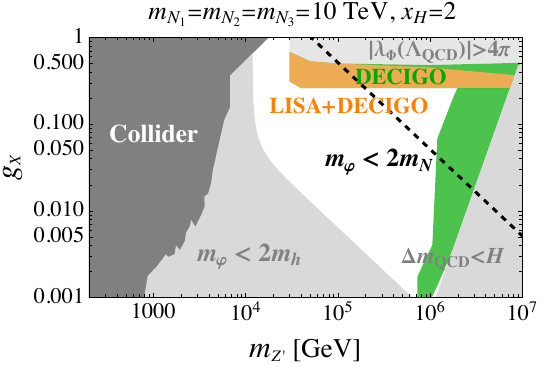}     
\caption{Parameter region explored by GW observatories and colliders for each benchmark point for $x_H=0$(top, B$-$L), $2$ (bottom), respectively for degenerate heavy neutrinos with masses 1 (10)TeV in the left (right) panel. In the colored regions, GWs can be detectable at each observatory. 
The dark gray region shows constraints by current collider searches from LEP and LHC. In the light gray regions, the perturbativity bound $|\lambda_{\Phi}|<4\pi$ or the requirements from Eq.~\eqref{eq:m_varphi_mh_mN} and Eq.~\eqref{eq:mQCD_Hubble} are not satisfied.}
\label{fig:2dim_mZp_gX-2}
\end{figure}

In Figs.~\ref{fig:2dim_mZp_gX-1} and \ref{fig:2dim_mZp_gX-2}, the parameter regions explored by colliders and each GW observation are shown for different values of $m_{N_{i}}$ and $x_{H}$. 
The dark gray region shows bounds from the measurements of the dilepton~\cite{ATLAS:2019erb,CMS:2019tbu} and dijet~\cite{CMS:2018mgb,ATLAS:2019bov} 
final states from $Z^\prime$ production at the LHC. 
In the orange and green regions with $g_{X} > 0.25$, the GWs can be produced by the conventional vacuum bubble dynamics in the cosmological first-order phase transition~\cite{Caprini:2024hue}. 
We use the fitting functions for the GW spectrum summarized in Ref.~\cite{Caprini:2024hue}. 
The typical frequency range is within the region explored by LISA and DECIGO. 
Furthermore, we have confirmed that several parameter regions with heavier $m_{Z'}$ may be tested by Cosmic Explorer~\cite{Reitze:2019iox} and Einstein telescope~\cite{Punturo:2010zz, ET:2025xjr}. 
For the green regions below $g_{X} \sim 0.25$, the tachyonic phase transition can produce the GWs that are detectable at DECIGO. 
In the lower right gray region, the condition for the classical rolling in Eq.~\eqref{eq:mQCD_Hubble} is not satisfied. 
For the lower left region, the condition $m_{\varphi} > 2 m_{h}$ is not satisfied. 
The black dashed line denotes the boundary below which the first requirement in Eq.~\eqref{eq:m_varphi_mh_mN} cannot be satisfied. 
Since the total decay width of the $\varphi$ field is changed below the boundary, the behavior of the green region is changed around the black dashed line. 
These figures indicate that we can explore wide parameters using GW observations even if the value of $U(1)_{X}$ gauge coupling is relatively small. 

\begin{figure}[t]
    \centering
    \includegraphics[width=0.65\linewidth]{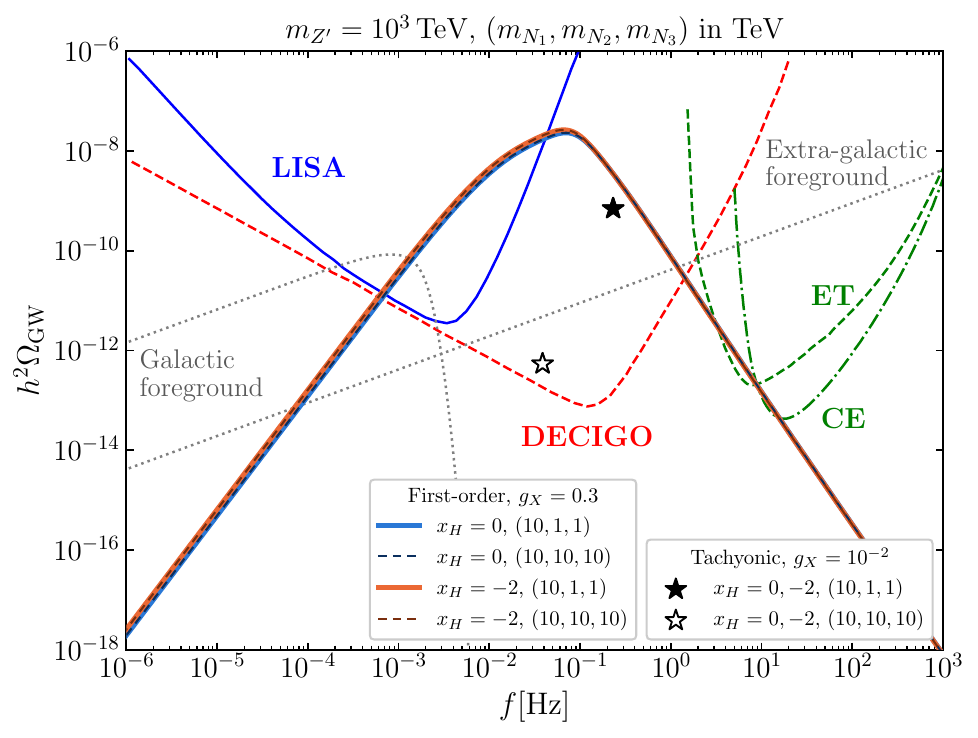}
    \caption{
    Predicted GWs in each benchmark point in Table~\ref{tab:BP_GW}.
    The sensitivity curves of LISA~\cite{Caprini:2024hue}, DECIGO~\cite{Kawamura:2006up,Yagi:2011wg}, Cosmic Explorer (CE)~\cite{Reitze:2019iox} and Einstein telescope (ET)~\cite{Punturo:2010zz, ET:2025xjr} are also shown. 
    The gray dotted lines denote the background signal from galactic foreground and extra-galactic foreground~\cite{Caprini:2024hue}. 
    }
    \label{fig:GW_mZp}
\end{figure}

In the preceding analyses, we have mainly focused on three RHNs with degenerate masses.
To illustrate the effects of relaxing this assumption, Tab.~\ref{tab:BP_GW} compares representative benchmarks with degenerate and nondegenerate RHN mass spectra at fixed $(m_{Z'},g_X,x_H)$.
For the first-order transitions considered here, the GW peak frequency and amplitude remain almost unchanged, as the transition dynamics is dominated by the $Z'$ boson contribution to the effective potential.
As shown in Fig.~\ref{fig:GW_mZp}, these benchmark points can be tested at LISA and DECIGO. 
For the tachyonic transitions, however, the RHN mass spectrum substantially affects both the peak frequency and amplitude.
In these benchmarks, the scalar decay $\varphi\to N_iN_i$ is kinematically allowed for the RHNs with masses of $1\,{\rm TeV}$ but forbidden for those with masses of $10\,{\rm TeV}$.
Replacing the lighter RHNs with heavier ones therefore reduces the total scalar decay width, changing the reheating temperature and the scale-factor ratio $r_{\rm RH}$.
The resulting changes in the redshift and dilution of the GWs shift the peak to lower frequencies and suppress its amplitude.
As a result, such benchmark points cannot be explored by DECIGO because their GW signals are below the background signal of extra-galactic foreground~\cite{Caprini:2024hue}. 
These results indicate that GWs from tachyonic phase transitions can be sensitive to the RHN mass spectrum through its impact on reheating.

\begin{table}[t]
\centering
\begin{tabular}{|ccccc|}
\hline
$(m_{N_1},m_{N_2},m_{N_3})\,[{\rm TeV}]$
& $(g_X,x_H)$
& Transition type
& $f_{\rm peak}\,[{\rm Hz}]$
& $\Omega_{\rm peak}h^2$
\\ \hline
$(10,1,1)$
& $(0.3,0)$
& First-order
& $0.071$
& $2.3\times10^{-8}$
\\
$(10,10,10)$
& $(0.3,0)$
& First-order
& $0.071$
& $2.3\times10^{-8}$
\\ \hline
$(10,1,1)$
& $(0.3,-2)$
& First-order
& $0.066$
& $2.6\times10^{-8}$
\\
$(10,10,10)$
& $(0.3,-2)$
& First-order
& $0.066$
& $2.6\times10^{-8}$
\\ \hline
$(10,1,1)$
& $(10^{-2},0)$
& Tachyonic
& $0.232$
& $7.0\times10^{-10}$
\\
$(10,10,10)$
& $(10^{-2},0)$
& Tachyonic
& $0.039$
& $5.4\times10^{-13}$
\\ \hline
$(10,1,1)$
& $(10^{-2},-2)$
& Tachyonic
& $0.232$
& $7.0\times10^{-10}$
\\
$(10,10,10)$
& $(10^{-2},-2)$
& Tachyonic
& $0.039$
& $5.4\times10^{-13}$
\\ \hline
\end{tabular}
\caption{
Transition types and present-day GW peak frequencies and amplitudes
for representative benchmarks with degenerate and nondegenerate RHN
mass spectra.
All benchmark points have $m_{Z'}=10^6\,{\rm GeV}$.
}
\label{tab:BP_GW}
\end{table}

\section{Conclusions}
\label{secVII}

We have explored the parameter region where the tachyonic phase transition occurs in the classically scale invariant general $U(1)_{X}$ extension of the SM with three heavy neutrinos of benchmark degenerate masses 1 TeV and 10 TeV, respectively.
In particular, if the mass of the new gauge boson $Z'$ and the $U(1)_{X}$ gauge coupling are within the range $10^{6}\,{\rm GeV} < m_{Z'} < 10^{7}\,{\rm GeV}$ and $0.001 < g_{X} < 0.25$, the GW signal from the tachyonic phase transition may be detectable at DECIGO.
On the other hand, we have found that the parameter region with $g_{X} > 0.25$ can be explored by observing GWs produced by the dynamics of vacuum bubbles nucleated by first-order phase transitions. The areas of the parameter spaces depend on the order of the RHN mass.
These parameter regions can be probed by future GW experiments like LISA and DECIGO.
Therefore, the parameter region with the heavy $Z'$ boson and the small $g_{X}$ where the collider exploration is challenging, can be tested via GW observations. 
This finding indicates that the complementarity of colliders and GW observations is useful in testing our model thoroughly.

\section*{Acknowledgments}

This work is supported by Grant-in-Aid for Early-Career Scientists 25K17398 (KH).\\

\appendix
\section{Renormalization group equations of the $U(1)_X$ model} 
\label{Sec:RGEs}
The RGEs for the SU(3)$_c$ and SU(2)$_L$ gauge couplings at the one-loop level are given by~\cite{Das:2016zue} 
\begin{align}
\begin{aligned}
&\frac{d g_3}{dt} = \frac{g_3^3}{(4\pi)^2} \left[ -7 \right] \,, \quad 
\frac{d g_2}{dt} = \frac{g_2^3}{(4\pi)^2} \left[ -\frac{19}{6} \right] \,, \quad 
\frac{d g_Y}{dt} = \frac{g_Y^3}{(4\pi)^2}\beta_Y \,, \\
&\frac{d g_{X}}{dt} = \frac{g_{X}^3}{(4\pi)^2}
\left(
\beta_X + 2\beta_{\rm mix} x_{\rm mix} + \beta_Y x_{\rm mix}^2
\right) \,, \quad 
\frac{d x_{\rm mix}}{dt} = 
\frac{ 2 g_Y^2}{(4\pi)^2} \left( \beta_Y x_{\rm mix} +\beta_{\rm mix} \right) \,,
\end{aligned}
\end{align}
where $\beta_Y=\frac{41}{6}, \beta_X = \frac16 \left( 72 + 64 x_H + 41 x_H^2\right)$ and $\beta_{\rm mix} = \frac16 \left( 32+41 x_H\right)$. 
For the RGEs for the Yukawa couplings at the one-loop level, we have 
\begin{align}
&\begin{aligned}
\frac{d y_t}{dt} = 
\frac{y_t}{(4\pi)^2} & \left[ \frac{9}{2}y_t^2 - 8g_3^2 - \frac{9}{4}g_2^2 - \frac{17}{12}g_Y^2 \right. \\ & \left. 
- g_X^2\left\{
\frac{17}{12} x_{\rm mix}^2+ \left(\frac{17}{6} x_H + \frac53 \right)x_{\rm mix} + \frac{17}{12} x_H^2 + \frac53 x_H + \frac23 \right\} \right] \,,
\end{aligned} \\
&\begin{aligned}
\frac{d Y_N}{dt}= \frac{Y_N}{(4\pi)^2} \left[ Y_N^\dagger Y_N + \frac{1}{2}\,{\rm Tr}\left(Y_N^\dagger Y_N \right) - 6 g_X^ 2 \right], 
\end{aligned}\\
&\begin{aligned}
\frac{d Y_\nu}{dt} = & \frac{Y_\nu}{(4\pi)^2}  \left[3 y_t^2 -\frac34 g_Y^2 -\frac94 g_2^2 -g_X^2\left\{ \frac34 x_{\rm mix}^2 + \left( 3+\frac32 x_H \right) x_{\rm mix} +\frac34 x_H^2 +3x_H+6
\right\} \right] \\
& +\frac{1}{(4\pi)^2}\left(
\frac{1}{2} Y_\nu Y_N^{} Y_N^\dagger + \frac32 Y_\nu Y_\nu^\dagger Y_\nu \right).
\label{Eq:RGE_Yukawa} 
\end{aligned}  
\end{align}
Finally, the RGEs for the scalar quartic couplings are given by 
\begin{align}
\begin{aligned}
\frac{d \lambda_H}{dt} = \frac{1}{(4\pi)^2} 
&\left[ \lambda_H \left\{ 24\lambda_H + 12y_t^2 + 4 Tr[Y_\nu^\dagger Y_\nu]- 9g_2^2 - 3g_Y^2 
- 3 g_X^2 \big( x_{\rm mix}+ x_H \big)^2 \right\}
\right.  \\
& + \left. 
\lambda_{\rm mix}^2 - 6 y_t^4 -2 {\rm Tr}[(Y_\nu^\dagger Y_\nu)^2]
+ \frac{9}{8}g_2^4 
+ \frac{3}{8}g_Y^4 + \frac38 g_X^4\big( x_{\rm mix}+ x_H \big)^4
\right. \\ & +\left.
\frac34 g_2^2 g_Y^2 + \frac34 g_2^2 g_X^2\big ( x_{\rm mix}+x_H \big)^2 
+ \frac34 g_Y^2 g_X^2 \big( x_{\rm mix}+ x_H \big)^2 \right] \,,
\end{aligned}
\end{align}

\begin{align}
&\begin{aligned}
\frac{d \lambda_\Phi}{dt} = \frac{1}{(4\pi)^2} 
 \left[ \lambda_\Phi \left\{ 20\lambda_\Phi + 2 {\rm Tr}[Y_N^\dagger Y_N] 
- 48  g_X^2 \right\} 
+ 2\lambda_{\rm mix}^2 - {\rm Tr}[(Y_N^\dagger Y_N^{})^2] 
+ 96 g_X^4 \right] \,,
\end{aligned} \\
&\begin{aligned}
\frac{d \lambda_{\rm mix}}{dt} 
= \frac{1}{(4\pi)^2} &
\left[ \lambda_{\rm mix} \left\{ 12\lambda_H + 8\lambda_\Phi + 4\lambda_{\rm mix} 
+ 6y_t^2 + {\rm Tr}[Y_N^\dagger Y_N^{}]  
\right. \right. \\ & \left.\left.
-\frac{9}{2} g_2^2 - \frac{3}{2} g_Y^2  -\frac32 g_X^2 \left(x_{\rm mix} + x_H \right)^2 - 24 g_X^2 \right\} 
\right. \\
&+ \left. 12 g_X^4 (x_{\rm mix}+x_{H})^2  - \frac12 {\rm Tr}[Y_N^\dagger Y_N^{} Y_\nu^\dagger Y_\nu]
\right].
\end{aligned}
\label{Eq:RGE_lambda}
\end{align}

\bibliographystyle{JHEP}
\bibliography{references}

\end{document}